\documentclass[fleqn,11pt]{article}

\usepackage{graphicx}
\usepackage{amsmath,amssymb,latexsym}
\usepackage{color}

\usepackage{epstopdf}

\usepackage{amsfonts,amssymb,cite}
\usepackage{graphicx}

\def\noi{\noindent}

\newcommand{\Title}[1]{\noi {{\Large\bf #1}}\\[1ex]}

\def\Aunames#1{\noi{\bf #1}}

\def\Addresses#1{\medskip\noi \protect
	\begin{description}\itemsep -3pt {\it #1} \end{description}}
\def\addr#1#2{\item[${}^{#1}$]{\it #2}}

\newcommand{\Abstract}[1]{\vskip 2mm \begin{center}
        \parbox{16.4cm}{\small\noi #1} \end{center}\medskip}

\def\email#1#2{\footnotetext[#1]{e-mail: #2}\addtocounter{footnote}{1}}

\def\nqq{\hspace*{-2em}}
\def\nhq{\hspace*{-0.5em}}

\def\cm{\hspace*{1cm}}
\def\inch{\hspace*{1in}}

\def\Jl#1#2{#1 {\bf #2},\ }

\def\ApJ#1 {\Jl{Astroph. J.}{#1}}
\def\CQG#1 {\Jl{Class. Quantum Grav.}{#1}}
\def\DAN#1 {\Jl{Dokl. AN SSSR}{#1}}
\def\GC#1 {\Jl{Grav. Cosmol.}{#1}}
\def\GRG#1 {\Jl{Gen. Rel. Grav.}{#1}}
\def\JETF#1 {\Jl{Zh. Eksp. Teor. Fiz.}{#1}}
\def\JETP#1 {\Jl{Sov. Phys. JETP}{#1}}
\def\JHEP#1 {\Jl{JHEP}{#1}}
\def\JMP#1 {\Jl{J. Math. Phys.}{#1}}
\def\NPB#1 {\Jl{Nucl. Phys. B}{#1}}
\def\NP#1 {\Jl{Nucl. Phys.}{#1}}
\def\PLA#1 {\Jl{Phys. Lett. A}{#1}}
\def\PLB#1 {\Jl{Phys. Lett. B}{#1}}
\def\PRD#1 {\Jl{Phys. Rev. D}{#1}}
\def\PRL#1 {\Jl{Phys. Rev. Lett.}{#1}}

\def\al{&\nhq}
\def\lal{&&\nqq {}}
\def\eq{Eq.\,}
\def\eqs{Eqs.\,}
\def\beq{\begin{equation}}
\def\eeq{\end{equation}}
\def\bear{\begin{eqnarray}}
\def\bearr{\begin{eqnarray} \lal}
\def\ear{\end{eqnarray}}
\def\earn{\nonumber \end{eqnarray}}
\def\nn{\nonumber\\ {}}

\def\nnn{\nonumber\\ \lal }

\def\yy{\\[5pt] {}}
\def\yyy{\\[5pt] \lal }
\def\eql{\al =\al}

\def\e{{\,\rm e}}
\def\d{\partial}

\def\im{\mathop{\rm Im}\nolimits}

\def\sign{\mathop{\rm sign}\nolimits}

\def\const{{\rm const}}
\def\eps{\varepsilon}

\def\then{\ \Rightarrow\ }

\def\da{\delta\alpha}
\def\db{\delta\beta}
\def\df{\delta\phi}
\def\dg{\delta\gamma}
\def\Veff{V_{\rm eff}}

\def\mN{_\mu^\nu}

\def\N{{\mathbb N}}

\def\rf{\eqref}
\def\eqn{\eq\ \eqref}

\def\asflat{asymptotically flat}

\def\bhs{black holes}
\def\wh{wormhole}
\def\whs{wormholes}
\def\sph{spherically symmetric}
\def\ssph{static, spherically symmetric}

\begin{document}
\twocolumn[
%\jnumber{3}{2017}

\vspace{10mm}

\Title{On wormholes with long throats and the stability problem}

\Aunames{K. A. Bronnikov$^{a,b,c,1}$ and P. A. Korolyov$^{a,2}$} 

\Addresses{
\addr a {Peoples' Friendship University of Russia (RUDN University), 
               ul. Miklukho-Maklaya 6, Moscow 117198, Russia}
\addr b {Center for Gravitation and Fundamental Metrology, VNIIMS,
               Ozyornaya ul. 46, Moscow 119361, Russia}
\addr c {National Research Nuclear University MEPhI
            (Moscow Engineering Physics Institute), Kashirskoe sh. 31, Moscow 115409, Russia}
	}

%\Rec{April 15, 2017}
	
\Abstract
{We construct explicit examples of globally regular static, spherically symmetric solutions of 
  general relativity with a phantom scalar field as the source of gravity, describing traversable
  wormholes with flat asymptotic regions on both sides of the throat as well as regular black holes,
  in particular, those called black universes. To explain why such phantom fields are not observed 
  under usual conditions, we invoke the concept of ``invisible ghosts,'' which means that the phantom 
  field decays quickly enough at infinity and is there too weak to be observed. This approach leads 
  to wormhole models in which the spherical radius is almost constant in some range of the radial
  coordinate near the throat, forming a "long throat". We discuss the peculiar features and difficulties 
  of the stability problem for such configurations. It is shown that the limiting case of a  "maximally 
  long throat" has the form of an unstable model with the Nariai metric. This allows us to
  conjecture that a long throat does not stabilize wormholes with a scalar source.
}  
     
]  %%%%%%%%%%%%%%%%%%%%%%%%%%%%%%%%%55
\email 1 {kb20@yandex.ru} 
\email 2 {korolyov.pavel@gmail.com}        

% ===============
\section{Introduction}

  Phantom field configurations have gained much interest since the discovery of the accelerated
  expansion of the Universe and its explanation in the framework of general relativity (GR) by the 
  existence of dark energy, a source of gravity of unknown nature which can violate 
  the standard energy conditions, such that the pressure to density ratio is $w = p/ \rho < -1$.
  Numerous observations lead to estimates of $w$ around $-1$, which corresponds to a cosmological
  constant, but values smaller than $-1$ are still admissible and even preferable for describing an
  increasing acceleration. Thus, one of the most recent estimates \cite{kb-Planck} reads 
  $w = -1.006 \pm  0.045$. 

  If, following many authors, we accept the existence of phantom (or exotic) matter as a 
  working hypothesis,  it is natural to expect that there exist its manifestations in local objects and
  phenomena. The simplest of them can be described by static, spherically symmetric solutions to the
  Einstein-scalar  equations where the scalar field has an unusual sign of kinetic energy (a phantom 
  scalar, by definition). In the case of a massless scalar, the corresponding solution \cite{kb-ber-lei} is a 
  phantom analog of Fisher's solution \cite{kb-fisher} for an ordinary minimally coupled 
  massless scalar field; the phantom-field solution \cite{kb-ber-lei} is sometimes called the anti-Fisher
  solution.

  It is also known that if one admits the existence of exotic matter, for example, in the 
  form of phantom scalar fields, there emerge such configurations of interest as 
  wormholes \cite{kb-k-73, kb-h_ellis, kb-extra-16} and regular black holes \cite{kb-bu1, kb-bu2}.

  Since no exotic matter or phantom fields have been detected under usual physical conditions, it is
  desirable to avoid the emergence of such fields in an asymptotic weak-field region, or at least
  to make them decay there rapidly enough. Thus, it was suggested \cite{kb-trap1} to use 
  a special kind of fields, named ``trapped ghosts,'' which have phantom properties only in 
  some restricted strong-field region and are usual. canonical in other parts of space. A variety 
  of solutions with such fields have been obtained, including regular 
  black holes, black universes and traversable wormholes \cite{kb-trap1,kb-trap2,kb-trap3}.
  In these models, the kinetic energy density smoothly passes zero at some scalar field value. 
  Such transition points create some problems with perturbation equations for these fields 
  \cite{kb-invis1}, which need a separate study.

  Another opportunity is to use phantom scalar fields rapidly decaying in weak-field regions. 
  In this paper we will consider a field of this kind and show that it leads to a \wh\ solution
  with a very slowly changing spherical radius in a neighborhood of the throat, so that
  it makes sense to speak of a \wh\ with a ``long throat''.\footnote
	{One can also mention an approach leading to configurations called time-independent
	\whs\ which, being constructed from the Schwarzschild-AdS metric, do not require 
         any exotic matter, see \cite{kb-fu-marolf} and references therein; they are related 
 	to the AdS-CFT correspondence and have peculiar expressions for the entropy. However, 
   	such objects are not wormholes in the terminology we are using: these are \bhs\ 
 	since they contain Killing horizons. These geometries also contain what can be called 
	anti-throats, i.e., local maxima of the spherical radius $r$.     
 } 

  Concerning the stability of such configurations, it turns out that a long throat causes substantial 
  technical difficulties because the effective potential for the most "dangerous" perturbations, 
  those preserving spherical symmetry, has such a pole on the throat that makes it very hard to 
  consider the perturbations of the throat radius. Meanwhile, it is this mode that makes unstable 
  many known wormhole and black universe solutions, as shown in \cite{kb-gonz,kb-sta1,kb-sta2}.
  These papers used the so-called S-transformation which regularizes the potential only if its pole 
  has the form $2/z^2 + O(1)$ (in properly chosen variables) but fails when dealing with other
  singularities that emerge in the case of a long throat. We postpone a detailed discussion of the 
  stability of long-throat \whs\ to future studies. Instead, we here consider, as a tentative problem, 
  the stability properties of a configuration with a constant spherical radius $r(x)$, which may be 
  called a "maximally long throat". It is not a \wh\ since there are no spatial asymptotic. 
  The corresponding static solution reduces to the well-known Nariai metric \cite{kb-nariai} 
  with a constant scalar field, and we explicitly show that it is unstable under linear perturbations. 
  Therefore, one can speculate that a slowly varying radius near a throat does not stabilize a 
  wormhole supported by a phantom scalar field.

  The paper is organized as follows. Section 2 presents the basic equations and some 
  general observations. In Section 3 we obtain examples of wormhole and regular black hole
  configurations. Section 4 discusses the linear stability problem for these solutions. Section 5 
  is devoted to stability study for a limiting configuration with a ``maximally long throat''.
  % Some concluding remarks are made in Section 6.
 
% ==========================
\section{Basic equations}

  We begin with the total action of GR with a scalar field source  
\beq         \label{act}
             S = \frac{1}{2} \int \sqrt{-g} d^4 x
	 \Big[R + 2 \eps  g^{\alpha \beta} \phi_{,\alpha} \phi_{,\beta} - 2V(\phi) \Big],
\eeq
  where $R$ is the scalar curvature, $\eps = 1$ for a normal scalar field with positive kinetic 
  energy, and $\eps = -1$ for a phantom scalar field;  $g = \det(g_{\mu \nu})$,
  and $V(\phi^a)$ is a self-interaction potential. The field equations may be written as
\bearr         \label{eq-phi}
	     2 \eps \nabla^{\mu} \nabla_{\mu} \phi + V_{\phi}= 0,
\yyy             \label{EE}
                R\mN = -2 \eps \phi_{ , \mu} \phi_{ , \nu} + \delta\mN V(\phi),
\ear
  and we use the units in which $8 \pi G = 1$ and $c = 1$.  

  Consider the general \ssph\ metric 
\beq                                   \label{ds}
       ds^2 = A(x) d t^2 - \frac{d x^2}{A(x)} - r^2(x)d \Omega^2,
\eeq
  in terms of the so-called quasiglobal radial coordinate $x$, such that $g_{00}g_{11}=-1$; 
  $d \Omega^2=(d \theta^2+\sin^2 \theta d \varphi^2)$ is the linear element on a unit sphere.
  Equations \rf{eq-phi} and \rf{EE} for the unknowns $\phi(x)$, $A(x)$ and $r(x)$ 
  take the form \cite{kb-BR}
\bearr        				 \label{eq-phi1}
              2(Ar^2 \phi')'  = r^2 \eps V_{\phi},
\yyy             			\label{Rtt}
           (A'r^2 )' = - 2r^2 V,
\yyy              			 \label{Rtt-Rxx}
            r''/r = - \eps \phi'^2 ,
\yyy                                     \label{Rtt-R22}
             A(r^2 )'' - r^2 A'' = 2,
\yyy              			\label{Gxx}
                -1 + A'r r' + A r'^2  = r^2 (\eps A \phi'^2  - V ),
\ear
  where the prime denotes $d/dx$, and $V_{\phi} = dV/d\phi$. Equation (\ref{eq-phi1}) 
  is the scalar field equation, the others are components of \rf{EE}, more specifically: 
  (\ref{Rtt}) is the component $R_t^t = \ldots$ , (\ref{Rtt-Rxx}) is the combination 
  $R_t^t-R_x^x = \ldots$, and (\ref{Rtt-R22}) is $R_{t}^{t}-R_{\theta}^{\theta}=\ldots$; lastly, 
  \eq (\ref{Gxx}) is the constraint equation for the Einstein tensor component $G_x^x = \ldots$. 
  Equations (\ref{eq-phi1}) and (\ref{Gxx}) follow from \eqs (\ref{Rtt})--(\ref{Rtt-R22}) 
  with a given  potential $V(\phi)$, so the latter form a determined set of equations for 
  $\phi(x), r(x), A(x)$. Equation (\ref{Rtt-R22}) can be integrated giving
\beq                        \label{B'(x)}
                B'(x) \equiv (A/r^2)' = (6m - 2x) / r^4, 
\eeq
  where $B(x) \equiv A/r^2$ and $m$ is an integration constant equal to the Schwarzschild 
  mass if the metric (\ref{ds}) is \asflat\ as $x\to \infty$ ($r \approx x$, $A = 1 - 2m/x + o(1/x)$).
  If there is a flat asymptotic as $x \to -\infty$, the Schwarzschild mass there is equal to 
  $-m$ ($r \approx |x|$, $A = 1 + 2m/|x| + o(1/x)$.

  Thus we have a general result: in {\it any\/} solution with two flat asymptotic 
  regions in the presence of {\it any\/} potential  $V(\phi)$ compatible with such behavior,  
  we inevitably have masses of opposite signs at the two infinities. Such solutions might in 
  principle represent \whs\ or regular \bhs. In the latter case, the function $B(x)$ would have 
  a minimum at which $B \leq 0$; it can be shown, however, that the existence of such a minimum 
  is incompatible with \eqn{Rtt-R22} \cite{kb-vac1}, therefore, if a solution to the above equations is 
  twice \asflat, it can only describe a \wh.  

% ====================================================
\section {Models with an invisible ghost and a long throat}

  Equations (\ref{eq-phi1})--(\ref{Gxx}) are hard to solve if a nonzero potential $V(\phi)$
  is specified. Instead, as in \cite{kb-BR,kb-bu1,kb-trap1,kb-trap2,kb-trap3}, we will use the inverse 
  problem method to find examples of interest: specifying the function $r(x)$, we can find all 
  other unknowns, including $V$, from the field equations: $A(x)$ is found from (\ref{B(x)}), 
  then $V(x)$ from (\ref{Rtt}), and $\phi(x)$ from (\ref{Rtt-Rxx}). 

  Our interest is in nonsingular configurations without a center, which can be wormholes or black
  universes \cite{kb-bu1}. Hence we assume that the range of $x$ is $x \in \mathbb{R}$, in which 
  both $A(x)$ and $r(x)$ are regular, $r > 0$ everywhere, and $r \to \infty$ at both ends. 
  We also require $r(x) \approx |x|$ as $x \to \pm \infty$, which is in agreement with possible
  flat, de Sitter or AdS asymptotic behaviors at large $r$. Thus there must be a minimum of 
  $r(x)$ (say, at $x=0$ without loss of generality), such that $r(0) > 0,$ $r'(0) = 0$, $r''(0) > 0$. 
  (It may happen that $r'' =0$ at a minimum of $r$ but then inevitably $r'' > 0$ in its neighborhood). 
  According to \eq (\ref{Rtt-Rxx}), this inevitably implies $\eps = -1$, a phantom field, which 
  is a manifestation of the well-known necessity of violating the Null Energy Condition (NEC)
  in wormhole and other regular spherical configurations in GR \cite{kb-BR, kb-thorne, kb-hoh-vis1}. 

  In our previous paper \cite{kb-invis1} we discussed a way to explain why such phantom fields 
  are not observed under usual conditions by using the concept of ``invisible ghosts,'' which means 
  that the phantom field decays quickly enough at infinity and is there too weak to be observed.
  We found in \cite{kb-invis1} some examples of \wh\ and regular black-hole solutions with and 
  without an electromagnetic field and two scalar fields, a long-range canonical one and a 
  comparatively short-range phantom one. Here we try to build similar models with a single 
  phantom field. To do so, that is, to achieve a sufficiently rapidly decaying scalar field energy
  density, we need rapidly decaying quantities $\phi'$ and hence, by \eq (\ref{Rtt-Rxx}), $r''/r$.
  We therefore replace the ansatz $r = a(1+x^2)^{1/2}$ used in some previous papers 
  (e.g., \cite{kb-bu1,kb-invis1}) with
\beq      \label{r(x)}
	 r(x) = a (1 + x^{2n})^{1/(2n)}, \cm n \in \N,
\eeq
  where $a > 0$ is an arbitrary constant equal to the throat radius. In what follows, we put 
  $a = 1$, which means that lengths are expressed in units of the throat radius; the quantities like 
  $B(x)$ and $V(x)$ with the dimension (length)$^{-2}$ are expressed in units of $a^{-2}$,
  while $A(x)$ and $\phi(x)$, being dimensionless, are insensitive to this assumption.

  The value $n=1$ returns us to the ansatz of our previous papers 
  \cite{kb-BR, kb-bu1, kb-invis1, kb-bu3}. With higher values of $n$, we obtain 
  a new feature of the space-time geometry: the spherical radius $r(x)$ is changing quite 
  slowly near the throat $x=0$, which enables us to call them {\it models with a long throat}.
  Indeed, with \rf{r(x)} at large $|x|$ we have $r''/r \approx (2n-1) x^{-2n-2}$,
  hence $\phi' \sim 1/x^{n+1}$, which at large enough $n$ conforms to the ``invisible ghost'' concept.
  On the other hand, at small $|x|$ we obtain $r(x) \approx 1 + x^{2n}/(2n)$, corresponding to 
  a ``long throat'' if $n >1$, see Fig.\,1a.     

  Let us put $m=0$, restricting ourselves to massless \whs. Then $B'(x)$ (\ref{B'(x)}) is 
  an odd function:
\beq                     \label{B'(x)1}
	 B'(x) = - \frac{2x}{(x^{2n} + 1)^{2/n}},
\eeq
  whose integration gives
\beq 		 \label{B(x)}
     B(x) =  -x^2  F \Big( \frac{1}{n}, \frac{2}{n};1 + \frac{1}{n}; -x^{2n} \Big) + B_0,
\eeq
  where $B_0$ is an integration constant, and $F(a, b; c, z)$ is the Gaussian 
  hypergeometric function. Assuming asymptotic flatness at large positive $x$, since $B = A/r^2$ 
  and $A \to 1$ at infinity, we require $B \to 0$ as $x \to\infty$ and thus fix $B_0$ as
\beq  			 \label{B0}
       B_0 = \lim\limits_{x\to \infty} x^2 F \Big( \frac{1}{n}, \frac{2}{n};1 + \frac{1}{n}; -x^{2n}\Big)
\eeq
  Plot of $B(x)$ and $A(x) = Br^2$ for $n=4$ are shown in Fig.\,1b,c. 
% ----------------------------------------------------------------------------- figure 1
\begin{figure*}
\centering
\includegraphics[width=5.5cm,height=5.5cm]{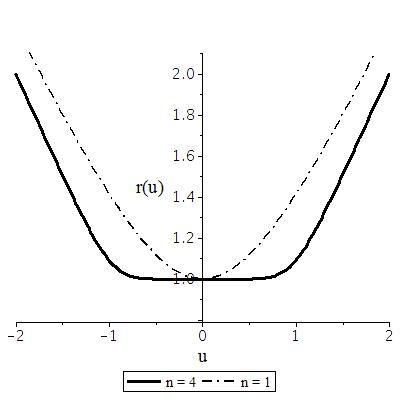}
\quad
\includegraphics[width=5.5cm,height=5.5cm]{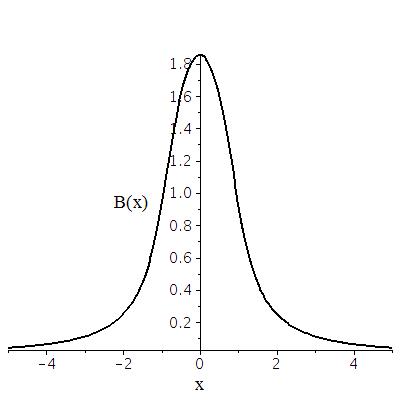}
\quad
\includegraphics[width=5.5cm,height=5.5cm]{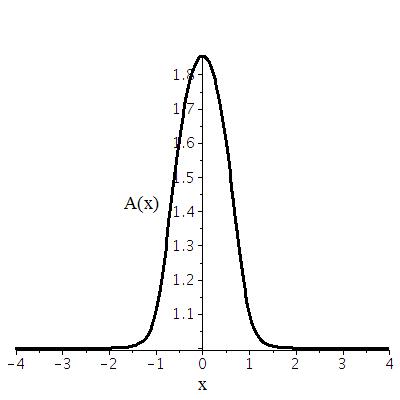}
\\
            a \inch\inch b\inch\inch c
\caption{\small 
     The long throat wormhole model: (a) $r(x)$ for $n=4$ (solid line, a long throat) and $n=1$ 
    (dot-dashed, a ``usual'' throat); (b) $B(x)$ and (c) $A(x)$, both for $n=4$.}
\end{figure*}
% -----------------------------------------------------------------------------
  We see that it is a twice asymptotically flat (M-M) wormhole (where ``M'' stands for Minkowski).
  Curiously, the behavior of $A(x)$ shows that there is a domain of repulsive gravity around the 
  throat. 

  Now we know the metric completely, and the remaining quantities $\phi(x)$ and $V(\phi(x))$ 
  can be easily found from \eqs (\ref{Rtt-Rxx}) and (\ref{Rtt}), respectively.
  The expression for the scalar field $\phi(x)$ for $n=4$ is (assuming $\phi(0)=0$)
\beq 		 \label{phix}
              \phi(x) =  \frac{\sqrt{7}}{4} (\sign x) \arctan (x^4),
\eeq
  see Fig.\,2a. For the potential $V(x)$ there is rather a cumbersome expression in terms 
  of hypergeometric functions, gamma functions and Legendre functions, and we will not 
  present it here. It is plotted in Fig.\,2b. Since $V(x)$ is an even function, the plot is restricted 
  to $x > 0$. 
% ----------------------------------------------------------------------------- figure 2
\begin{figure*}
\centering
\includegraphics[width=5.5cm,height=5.5cm]{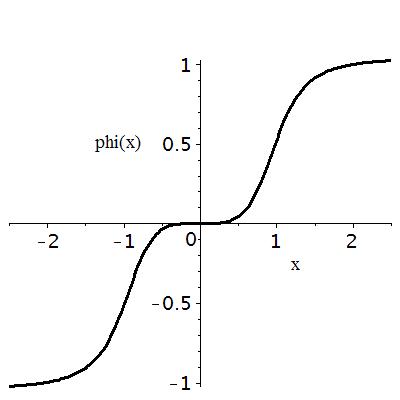}
\qquad
\includegraphics[width=5.5cm,height=5.5cm]{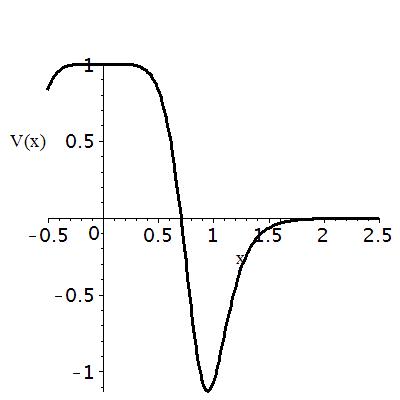}
\\
            a \inch\inch b
\caption{\small The scalar field $\phi(x)$ (a) and the potential $V(x)$ (b) for $n=4$.}
\end{figure*}
% -----------------------------------------------------------------------------

  More complicated models with various global structures emerge if there is a nonzero 
  Schwarzschild mass $m$ or/and, in addition to the scalar field, there is an electromagnetic 
  field with the corresponding electric or magnetic charge. Examples of such solutions, which 
  include M-M, M-dS (de-Sitter), M-AdS (anti de-Sitter) wormholes and regular black holes with 
  up to four horizons, are presented in \cite{kb-bu1,kb-bu3,kb-trap2,kb-trap3,kb-invis1}. Since the global 
  qualitative features of the present field system are similar to those described there, the same kinds
  of regular solutions can be obtained in the present case using the same methods. We will not 
  analyze them here, but, instead, briefly discuss the stability problem in the case of long throats.

% ====================================================
\section{The stability problem in the presence of a long throat}

  As is the case for any static configurations, a problem of importance is their stability or, more 
  generally, the behavior of their perturbations. For instance, certain systems can exist for quite a 
  long time even if they are unstable but decay very slowly. On the other hand, the evolution of 
  unstable systems can lead to many phenomena of interest, from structure formation in the 
  Universe to Supernova explosions.

  Small (linear) radial perturbations of scalar-vacuum configurations with various potentials 
  $V(\phi)$, including wormholes and regular black holes, have been studied in many papers, e.g., 
  \cite{kb-br-hod,kb-gonz,kb-sta1,kb-sta2}. The perturbations preserving spherical symmetry are, on 
  one hand, the simplest, and, on the other, the most destructive ones, and have been shown to
  lead to instabilities in many configurations with self-gravitating scalar fields, including 
  Fisher's solution \cite{kb-br-hod} and solutions with throats including wormholes 
  \cite{kb-gonz,kb-sta1,kb-sta2}. Let us briefly recall the relevant formalism, assuming that a certain 
  static solution is known (non necessarily one of those described here) and considering its 
  time-dependent perturbations.

  We begin with the general spherically symmetric metric\footnote
	{In the whole paper we are using the following notations for different radial coordinates:\\
	\cm      $u$ --- a general notation,\\
	\cm  	  $x$ --- quasiglobal, such that $\alpha = -\gamma$,\\
	\cm 	  $z$ --- ``tortoise", such that $\alpha = \gamma$.}
\beq                \label{ds2}
          ds^2 = e^{2\gamma}dt^2  - e^{2\alpha}du^2 - e^{2\beta}d\Omega^2 
\eeq
  where $\alpha$, $\beta$, $\gamma$ are functions of the radial coordinate $u$ (not necessarily the 
  quasiglobal coordinate $x$ used before) and the time $t$. 

  Preserving only linear terms with respect to time derivatives, we can write the field 
  equations \rf{eq-phi} and \rf{EE} as follows:
\beq                \label{phi..}
        \e^{-2\gamma}\ddot\phi - \e^{-2\alpha} [\phi'' + \phi'(\gamma'+2\beta'-\alpha')]
     		 =\eps  V_\phi/2;
\eeq
\vspace{-8mm}
\bear       \nhq                                  \label{R00}
     R^t_t \eql  \e^{-2\gamma}(\ddot\alpha + 2\ddot\beta) 
		- \e^{-2\alpha}\big[\gamma'' 
\nnn \quad\
		+\gamma' (\gamma'-\alpha'+2\beta')\big] = V(\phi),
\yy          \nhq                                \label{R11}
     R^1_1 \eql  \e^{-2\gamma}\ddot\alpha  
		- \e^{-2\alpha}\big[\gamma''+2\beta''+\gamma'{}^2 +2\beta'{}^2  
\nnn \ \ 
	- \alpha'(\gamma'+2\beta')\big]  = +2\eps \e^{-2\alpha}\phi'{}^2 + V(\phi),
\yy          \nhq                                            \label{R22}
     R^2_2 \eql R^3_3 = \e^{-2\beta}
          +\e^{-2\gamma}\ddot\beta -\e^{-2\alpha}\big[\beta''
\nnn \quad\
	+\beta'(\gamma'-\alpha'+2\beta')\big] = V(\phi),
\yy          \nhq                                                                         \label{R01}
     R_{01}\eql 2\big[\dot\beta' + \dot{\beta}\beta'
                 -\dot{\alpha}\beta'-\dot{\beta}\gamma'\big] = - 2\eps {\dot\phi}\phi',
\ear
  where dots denote $\d/\d t$, primes denote $\d/\d u$, and $V_\phi = dV/d\phi$.

  We assume that we know a static, purely $u$-dependent solution of this set of equations,
  and then consider small time-dependent deviations from it, denoted by adding the symbol 
  $\delta$, so that
\bearr
     	\phi(u,t) = \phi(u) + \df(u,t), 
\nnn  
	\alpha(u,t) = \alpha(u) + \da(u,t),
\earn 
  and so on, where $\phi(u), \alpha(u), \ldots$ are static solutions, while $\df(u,t), \da(u,t), \ldots$
  are small deviations. 

  Let us note that we have two independent forms of arbitrariness in \eqs \rf{phi..}--\rf{R01}: 
  the choice of a radial coordinate $u$ (used in the static solution and possibly fixed by a 
  relation between  $\alpha(u)$, $\beta(u)$, $\gamma(u)$) and the choice of a perturbation 
  gauge, corresponding to the choice of a reference frame in perturbed space-time (fixed by 
  a relation for the perturbations). The above equations \rf{phi..}--\rf{R01} are written in a 
  universal form, without coordinate or gauge fixing.

  In the \sph\ Einstein-scalar field system there is actually only one dynamic degree of freedom.
  Accordingly, in \eqs \rf{phi..}--\rf{R01} with respect to the perturbations, one can exclude 
  all unknown except $\df$ and, after passing over from an arbitrary coordinate $u$ to the 
  so-called tortoise radial coordinate $z$ defined by 
\beq                           \label{to_z}
       du/dz = \e^{\gamma-\alpha}
\eeq
  and separating the variables by the assumption 
\beq
       \df (z,t) = \e^{-\beta (u)} Y(z) \e^{i\omega t}, \qquad \omega = \const,                 
\eeq
  reduce the stability problem to a boundary-value problem for the Schr\"odinger-like equation 
\beq                                                        \label{Schr}
          d^2 Y/dz^2 + [\omega^2 - \Veff(z)] Y =0,
\eeq
  with certain physically motivated boundary conditions. The effective potential $\Veff$ has the form
\bearr                                                            \label{Veff}
          \Veff (z) =   \e^{2\gamma}\biggl[
	\frac{2\eps \phi'^2}{\beta'^2}  (V - \e^{-2\beta})
         	 + \frac{2V'}{\beta'} + \frac{\eps V_{\phi\phi}}{2}\biggr]
\nnn \cm
         + \e^{2\gamma-2\alpha}\Bigl[\beta'' + \beta' (\beta'+\gamma'-\alpha')\Bigr].
\ear
  where the prime, as before, denotes $d/d u$, and $V=V(\phi)$ is the original scalar field potential. 
  If there is a nontrivial solution to \eqn{Schr} with $\im \omega < 0$ satisfying physically reasonable 
  conditions at the ends of the range of $z$, then the static system is unstable since $\df$ can exponentially 
  grow with time. Otherwise our static system is stable in the linear approximation.

  A more detailed derivation can be found, e.g., in \cite{kb-sta1,kb-sta2,kb-BR,kb-lbook}.
  It should be noted that \eqn{Schr}, being derived using the convenient gauge $\db =0$,
  is still gauge-invariant since $Y$ can be shown to be a gauge-invariant quantity while $\Veff$ contains 
  quantities from the static background solution, written in terms of an arbitrary radial coordinate $u$.

  Now, to apply this general formalism to a static solution from the previous section, we must find 
  $\Veff(z)$, identifying $u$ with our quasiglobal coordinate $x$, and substitute 
  $\e^{2\gamma} = \e^{-2\alpha} = A(x)$, $\e^\beta = r(x)$. It is easy to see that at flat spatial 
  infinity the coordinates $z$ and $x$ almost coincide ($x \approx z$ for $x\to \pm\infty$), and the 
  natural boundary condition $\df(\pm\infty) =0$ leads to $Y(\pm\infty) = 0$.

  The main problem with this stability study is connected with the shape of the effective potential
   \rf{Veff}, which has a singularity at the throat due to $\beta' = r'/r  = 0$. More precisely, if 
  the throat is located at $z = 0$ and, as $z \to 0$, the radius $r$ behaves as 
  $r \approx r_0 + O(z^{2n})$ (as is the case under the ansatz \rf{r(x)}),\footnote
	   {Since, by \rf{to_z}, the coordinates $x$ and $z$ are related by $dz = dx/A(x)$, and $A(x)$ 
	     is finite at the throat, the function $r(z)$ qualitatively behaves in the same way as $r(x)$.}
  we have at small $z$
\bear                 \label{z0}
           \Veff (z) \eql 2\biggl[ \Big(\frac{\beta''}{\beta'}\Big)^2 - \frac{\beta'''}{\beta'}\biggr] + O(1)
\nn
                         \eql \frac{2(2n{-}1)}{z^2} + O(1),     
\ear
  where, as before, $\beta = \ln r$, while the prime here denotes $d/d z$. Recall that models with 
  ordinary throats correspond to $n=1$ and those with long throats to $n > 1$.

  As described, e.g., in \cite{kb-sta1, kb-BR, kb-lbook}, the potential $\Veff$ for $n=1$ can be 
  regularized using  the ``S-transformation,'' which is a special kind of substitution in \eqn{Schr}: 
  the transformed equation has no singularities, and solutions to the corresponding boundary-value 
  problem describe regular perturbations of the scalar field and the metric. This method was used 
  to prove the instability of  anti-Fisher \whs\ \cite{kb-gonz} and other configurations with throats 
  \cite{kb-sta1,kb-sta2}. The instability is caused by the perturbation mode in which the throat 
  radius changes with time.

  However, according to \cite{kb-sta1, kb-BR, kb-lbook}, a necessary condition for the S-transformation
  to remove the singularity in $\Veff$ is that $\Veff = 2/z^2 + O(1)$, which, by \eqn{z0}, is only 
  true for a ``usual'' throat, $n=1$. Therefore, a stability study of  long-throat wormholes faces 
  a serious technical difficulty, and to obtain an idea of whether or not a long throat can stabilize a
  wormhole, in the next section we will consider a simple model with 
  $r=\const$, which can be called a maximally long throat.   

% ======================================
\section{Instability of a maximally long throat}

  Consider a static solution to \eqs \rf{phi..}--\rf{R22} under the condition 
  $r\equiv \e^\beta = r_0 = \const$. Thus it will be not a wormhole but rather a ``pure throat''. 

  Choosing the coordinate $u=z$ such that $\alpha=\gamma$, from \rf{phi..}--\rf{R22} we easily obtain
\bearr                  \label{bsol1}
     \phi = \phi_0 = \const,  \qquad V = 1/r_0^2 =\const , 
\yyy			 \label{bsol2}
       ds^2 = \frac {k^2 r_0^2 (dt^2 - dz^2)}{\cosh^2(k z)} - r_0^2 d\Omega^2,
\ear
  where $k > 0$ is an integration constant, which, without loss of generality, can be put equal to $1/r_0$
  by choosing scales along the $t$ and $z$ axes. From the scalar field equation it also follows $V_\phi=0$
  at the value of $\phi$ corresponding to the solution. The metric can also be written in terms of
  the quasiglobal coordinate $x = r_0\tanh (z/r_0)$:
\bear                  \label{ds3}
             ds^2 \eql \frac{dt^2 - dz^2}{\cosh^2(z/r_0)} - r_0^2 d\Omega^2
\nn 
		\eql  (1-x^2/r_0^2) dt^2 - \frac{dx^2}{1-x^2/r_0^2} - r_0^2 d\Omega^2. 
\ear
  It is the well-known Nariai solution, a vacuum solution of GR with 
  the cosmological constant equal to $1/r_0^2$ \cite{kb-nariai, kb-exact}. We see that it can also be 
  interpreted as a special solution to GR with a scalar field source. The values $x = \pm r_0$,
  or equivalently $z = \pm\infty$, are horizons, and the static region between them represents what
  may be called a maximally long throat; its full proper length along the $z$ direction is equal 
  to $\pi r_0$.   

  Consider small time-dependent perturbations of this solution, using the coordinate $z$. 
  Then, without any assumption on the perturbation gauge, \eqs \rf{phi..}, \rf{R22}
  and \rf{R01} lead to decoupled equations for the unknowns $\df(z,t)$ and $\db(z,t)$:
\bearr     \label{phi-3}
                     \delta\ddot\phi  -  \df'' + \frac\eps 2\, e^{2\gamma}V_{\phi\phi}\df = 0,
\yyy                             \label{r01}
                     \delta {\dot \beta}' = \gamma' \delta {\dot \beta},
\yyy                             \label{r22}
                    \delta \ddot\beta - \db'' - \frac{2e^{2\gamma}}{r^2 _0}\db = 0,
\ear
  where the prime stands for $\d/\d z$ and the dot for $\d/\d t$. The existence of two independent 
  dynamic degrees of freedom for radial perturbations instead of a single, scalar one in the general 
  \wh\ case is connected with the absence of a Birkhoff-like theorem  in this case, see a 
  detailed discussion in \cite{kb-birk} and references therein. Perturbations of the spherical radius 
  actually behave in our case as one more scalar field in the 2D space-time parametrized by $t$ 
  and $z$. 

  Equation (\ref{r01}) admits integration in time, after which, neglecting an arbitrary function of $z$
  (a static perturbation), we get
\beq                     \label{db} 
                       \db' = \gamma' \db \ \then \ \db = v(t) \e^\gamma,
\eeq
  where $v(t)$ is an arbitrary function which may be further determined from \eqn{r22}. Substituting 
  \rf{db} into \rf{r22} and taking into account that, according to \rf{ds3}, 
   $\e^\gamma = 1/\cosh(kz)$, and $k = 1/r_0$, we obtain  $\ddot v - k^2 v =0$, so that finally
\beq         \label{sol-db}
                \db = v(t) \e^\gamma = \frac {c_1 \e^{kt} + c_2 \e^{-kt}}{\cosh(kz)},
\eeq 
  with arbitrary constants $c_1$ and $c_2$. The existence of the growth factor $\e^{kt}$ means that
  the background maximally long throat solution with the Nariai metric is unstable. 

  Since we did not so far use any perturbation gauge, there still remains a doubt that the behavior 
  of $\db(x,t)$ may be a pure gauge and may be removed by a $t$-dependent coordinate 
  transformation. To make sure that this instability is real, let us reconsider 
  \eqs \rf{phi..}--\rf{R01}, now using a manifestly admissible gauge \cite{kb-lbook} $\db = 0$. 
  Then \eqn{phi-3} preserves its form, but \eqs \rf{r22} and \rf{r01} become trivial, and we 
  should use the remaining equations \rf{R00} and \rf{R11}, which take the form
\bearr 
	\delta {\ddot\alpha} - \dg'' = 2 \e^{2\gamma}\da/r_0^2,
\nnn
        	\delta {\ddot\alpha} - \dg'' - \gamma'(\da' - \dg') = 2 \e^{2\gamma}\da/r_0^2.
\ear
  Their difference gives $\dg' = \da'$, and we arrive at a wave equation for $\da$
\beq
	\delta {\ddot\alpha} - \da'' = 2 \e^{2\gamma}\da/r_0^2,
\eeq
  the same as we previously had for $\db$, \eqn{r22}. Consequently, it has the solution similar
  to \rf{sol-db}
\beq
	\da = \frac {c_1 \e^{kt} + c_2 \e^{-kt}}{\cosh(kz)},
\eeq
  and we can confidently conclude that the maximally long throat solution is unstable. 
  
  This actually confirms the old result of \cite{kb-fujii} on the instability of Nariai's solution since 
  the scalar field contribution here reduces to supplying a cosmological constant $\Lambda = 1/r_0^2$. 
  The scalar field perturbations have, in the linear approximation, their own dynamics described 
  by \eqn{phi-3} and can be stable or unstable depending on the sign of $V_{\phi\phi}$.

  The instability of the limiting model of a ``maximally long throat'' allows us to conjecture that a slowly 
  varying radius near a throat of a more general \wh\ supported by a phantom scalar field
  does not stabilize it as compared with models with a ``usual'' throat. We hope to verify this 
  conjecture in our future work. Of interest are also the stability properties of other solution 
  with long throats, such as \whs\ with an AdS asymptotic at the far end as well as black universes
  with a horizon and an asymptotically de Sitter expansion beyond it, like those described in 
  \cite{kb-bu1,kb-bu2,kb-bu3}. 

\subsection*{Acknowledgments}

  We thank Sergei Bolokhov and Milena Skvortsova for numerous helpful discussions. 
  The work of KB was partly performed within the framework of the Center FRPP 
  supported by MEPhI Academic Excellence Project (contract No. 02.a03.21.0005, 27.08.2013).
  This work was also funded by the RUDN University Program 5-100 and by 
  RFBR grant 16-02-00602.

\end{document}